\documentclass[aps,twocolumn]{revtex4}
\usepackage{graphics}
\usepackage{dcolumn}
\begin{document}
\def\be{\begin{equation}}
\def\ee{\end{equation}}
\def\lag{\langle}
\def\rag{\rangle}
\title{Hamiltonian model for multidimensional epistasis} 
\author{Seher \"Oz\c celik$^1$ and
Ay\c se Erzan$^{1,2}$}
\affiliation{$^1$  Department of Physics, Faculty of  Sciences
and
Letters\\
Istanbul Technical University, Maslak 80626, Istanbul, Turkey }
\affiliation{$^2$  G\"ursey Institute, P. O. Box 6, \c
Cengelk\"oy 81220, Istanbul, Turkey}
\date{\today}
\begin{abstract}
We propose and solve a Hamiltonian model for multidimensional epistastatic interactions between beneficial mutations.  The model is able to give rise either to a phase transition between two equilibrium states, without any coexistence, or exhibits a state where hybrid species can coexist, with gradual passage from one wild type to another. The transition takes place 
as a function of the {\it tolerance} of the environment, which we define as the amount of noise in the system.  


\end{abstract}
\maketitle
%\begin{twocolumn}
%Introduction
\section{Introduction}

Evolution takes place via natural selection, whereby random mutations
which have a salutary effect on the fitness (survival probability and/or
reproductive capability) of the individual persist in the population and
lead to new variants; other, neutral mutations may simply be carried along
since they do not affect the well being of the individual. Deleterious
mutations usually affect the organism adversely, and the accumulation of
too many will reduce the fitness drastically. Each ``species'' actually
consists of a more or less narrow distribution in the phase space of all
possible genetic states, and this distribution may shift, in response to
environmental pressure, along definite evolutionary routes.~\cite{Smith}

An interesting problem is to explain the rather fast rates at which 
populations seem to be able to adapt to changing environments, which 
suggest that the fitness does not depend in a simple additive way on the 
effects of each independent mutation, but that there is a nonlinear 
relationship, or epistasis~\cite{Book}, between the effects of point 
mutations 
determining the fitness function.  In fact, one may surmise that evolution is not a Markovian game, but that the fitness depends upon the history of the successive mutations, in other words, it is a function of the path taken in genomic space.~\cite{Kondrashov}  
Thus, for a mutation leading to the development of fingers to be beneficial, say, one must already have had a mutation leading to the formation of limbs.

Posed in this way, this problem seems to demand an analysis that is intrinsically dynamical. Yet,  it actually lends itself to a treatment in terms of statistical equilibria, with the appropriate choice of a fitness function.  
In this paper we aim to provide such a fitness function, and solve the resulting model for possibly coexisting equilibrium phases indicating different species.

\section{The Model}

Since Eigen first introduced the quasi-species model~\cite{Eigen} there
has been a huge amount of both analytical and numerical work on bitstring
models of genetic evolution~\cite{Stauffer,Oliveira}, where the genotype 
of an
individual is represented by a string of Boolean variables $\sigma_i$,
$i=1,\ldots N$. If one takes the wild type, or the initial genotype, to
consist of a string of 0's, each point mutation is indicated by flipping
the bit respresenting a given gene, from 0 to 1. The number of mutations
$m$ is then the number of 1's on the whole string, i.e.,
$m=\sum_i\sigma_i$. The fitness is usually taken simply to be a function,
albeit nonlinear, of $m$.

Clearly, each $i$'th variable can be considered as an independent
direction in phase space, so that evolution takes place in an $N$
dimensional space, where $N$ is the length of the genome. The genotype is
a vertex on an $N$-dimensional unit hypercube, and if only single flips
from 0 to 1 are allowed at a time, the path of evolution is a 
random walk on the
edges of this hypercube. One possible way in which a vector variable can
be introduced is to consider the whole vector ${\bf V}\equiv \{\sigma_i\}$
as the argument of the fitness function. Since the position of each gene
on this particular string can be assigned with some arbitrariness, one may
then demand that the fitness is only increased relative to the wild type
if the bits that flip to 1 occur sequentially. \cite{Kondrashov}. Thus,
$(0,0,\ldots)$, $(1,0,\ldots)$$(1,1,0\ldots)$ are in increasing order of
fitness while $(0,1,\ldots)$ is less fit.

This demands that we introduce a cost function $H$ which depends on the state ${\bf V}$, and we have chosen the fitness $f$ to decrease exponentially with this cost function, viz.,
\begin{equation}
f\propto e^{-\beta H} \label{Boltz}
\end{equation}
where $\beta$ is a measure of how effective the cost function is, in
affecting the fitness. The fitness function $f$ can be identified as the
Boltzmann factor in an equilibrium statistical model with the Hamiltonian
$H$, at constant inverse ``temperature" $\beta^{-1}$. Temperature may be
seen as the amount of randomness, or disorder in the system, competing
with the cost function in determining the fitness.  The higher the
temperature, or randomness, the weaker will be the effect of the cost
function in determining the state of the system. Therefore we define
\be T\equiv\beta^{-1}\ee
as the {\it tolerance} in the system.

For the cost function we will borrow a Hamiltonian introduced by Bakk at 
al.~\cite{Bakk}, in the context of protein folding, where it is of importance that the folding events take place in a prescribed order. Thus, 
\begin{equation}
H=- \lambda J\sum_{m=1}^N \prod_{i=1}^m \sigma_i - (1-\lambda) J
\prod_{i=1}^N \sigma_i\;\;\;. \label{Ham}
\end{equation}
It can be seen that for  $\lambda = 0$, the only state which is favorable 
is that with all $\sigma_i=1$, whereas for $0 < \lambda\le 1$, all states 
with an uninterrupted initial sequence of $1$s of arbitrary length $m$ lead to improved fitness.  Here $J$ is a measure of the strength of the interaction between the states of each of the sites (alleles), $\sigma_i$.  Clearly, $\beta$ and $J$  will always occur together in this model, in the product $\beta J$, and we may simply absorb $J$ into the definition of $\beta$.
         
The fitness $f$ may be normalized to take values between $(0,1)$ if we devide the expression in Eq.(\ref{Boltz})
by the sum over all states ${\bf V}$, namely, 
\begin{equation}
Z\equiv \sum_{\{\sigma_i\}} e^{-\beta H[\{\sigma_i\}]}\;\;\;.
\end{equation}
This sum may be performed exactly, to give,
\begin{equation}
Z= {2^N - e^{\beta \lambda   N} \over 2- e^{\beta \lambda }} + 
e^{[\lambda (N-1)+1]\beta  } \;\;\;\;.\label{Z}
\end{equation}
Using this result we may compute the expectation values (average values) of the quantities $\psi_m\equiv  \prod_{i=1}^m \sigma_i$, which we shall call, 
$ \Psi_m=\langle \psi_m
\rangle$, for $m=1, \ldots, N$.  Clearly, $\Psi_m$ is the probability that 
in equilibrium, at least $m$ initial loci on the genotype have switched to 1.  One finds,
\begin{equation}
\Psi_m= {1 \over Z} 
\left[ {2^{N-m} e^{m \beta \lambda } - e^{N\beta \lambda } 
\over 2-e^{\beta \lambda }} 
+ e^{\beta  [(N-1) \lambda + 1]}\right]\;\;\;.\label{Psi}
\end{equation}
Clearly, as $\beta \to 0$, $\Psi_m \to (1/2)^m$, so that it is convenient  to define the order parameters 
\be
\Phi_m = {\Psi_m - (1/2)^m \over 1-(1/2)^m}\;\;\;, \label{Phi}\ee
which take values in the interval $(0,1)$.  In Fig. 1, we present the results of a numerical evaluation of Eq.(\ref{Psi}), for $\lambda=0$, and in Fig.2, for $\lambda=1$, as a function of $x\equiv T/J$, which is the ratio of   the tolerance in the system to the strength of the epistatic interactions.

We find that for $\lambda=0$, there is a sharp  transition for large $N$, at 
$x > x_t$, below which the genotype is completely ordered, with all 
$\sigma_i=1$, while for $x > x_t$, the whole population is in the the 
state 
with all $\sigma_i=0$. From an inspection of Eqs.(\ref{Z},\ref{Phi}), one 
sees that $x_t=(N \ln 2)^{-1}$. Thus there are only two possible species in this case, with no coexistence between them.  However, for $N\to \infty$, the threshold itself goes to zero. (This can be mended if the strength of the second term in Eq.(\ref{Ham}) is chosen to be $NJ$ rather than $J$.)

For $\lambda=1$, it can be seen that the sharp phase transition is no
longer present (the nonzero value of $\lambda$ has an effect similar to
turning on a magnetic field in a magnetic phase transition).  For large
$x$, all the $\Phi_m$ decay exponentially, as $\sim e^{-m/x}$.  However,
there exist effective thresholds $x_m$, for $m > 1$, below which there is
a nonzero probability of encountering individuals with $m$ initial alleles
switched to 1.  This signifies that at any given $x_{m+1}< x < x_m$, there
is coexistence between $m$ hybrid species, with the first $n\le m$ alleles
in the mutated state.  The probability of encountering an individual with
$n\le m$ seqentially mutated alleles is in fact precisely $\Phi_m$. We see
that $\Phi_N\approx 0$ for $x > x_N$, with $x_N \sim {1/\ln2}$.

To further elucidate the meaning of ``tolerance,'' we may compute the
relative variances $v_m$, where

\be
v^2_m \equiv \langle (\psi_m  - \Psi_m)^2 \rangle /\Psi_m^2\;\;\;.
\ee
It is trivial to note that $\psi_m^2 = \psi_m$, so that $v_m^2 =
(1-\Psi_m)/\Psi_m$.  Then it is straightforward enough to get,
\be v_m^2 = {1- \left(e^{\beta \lambda}/2\right)^m \over 
\left(e^{\beta \lambda}/2\right)^m + \left(e^{\beta \lambda}/2\right)^N\left[2e^{\beta(\lambda-1)}-e^\beta -1\right]}\;\;\;.
\ee
One may see from here that the behaviour of the system is determined by 
the critical value of $\beta \lambda$ at $\ln 2$, and moreover, that the 
variance (or the departure from the ordered phase) also depends on whether $m$ is small or of the order of $N$, as can also be seen clearly from Fig. 2.

\section{Conclusion} 
In summary, we have presented a Hamiltonian model of
multidimensional epistasis which weights only certain paths in genotype
space as being favorable.  The model has tunable strength ($J$) of
interactions between different genes, which can be absorbed into an
overall parameter ($\beta$) which determines how tolerant the environment
is to deviations from the wildtype, as well as a parameter which decides
whether coexistence between hybrid individuals will or will not be
allowed.  The model exhibits a transition between two pure types as a
function of $\beta$ for $\lambda=0$.  For $\lambda \ne 0$, low tolerances
$T=\beta^{-1}$ give rise to the appearance of hybrid types, in case a
given series of mutations increases the fitness.

{\bf Acknowledgements}

We are grateful to G\"une\c s S\"oyler for many useful discussions.  AE
acknowledge partial support from the Turkish Academy of Sciences.

%\end{document}

{\bf Figure Captions}

Fig. 1. The order parameters $\Phi_m$ for $\lambda=0$ all differ from 
zero at the same transition point. Here the length of the genome is 100. 
There are no hybrid species. 

Fig. 2. The order parameters $\Phi_m$ for $\lambda=1$, with $N=100$.  
There is a set of $N$ effective thresholds, below which hybrid species
arise, with $m$ sequentially mutated alleles.

%%%%%For some reason I could not figure out, the figures did not show up 
%%%%%when included like this; so I took this part out and sent them 
%%%%%seperately. 
%\begin{figure}
%\leavevmode
%\rotatebox{0}
%{\resizebox{6cm}{5cm}{\includegraphics*{seherfig1.eps}}}
%\caption{The order parameters $\Phi_m$ for $\lambda=0$ all differ from 
%zero at the same transition point. Here the length of the genome is 100. 
%There are no hybrid species. }
%\end{figure}   

%\begin{figure} %\leavevmode
%\rotatebox{0}{\resizebox{6cm}{5cm}{\includegraphics*{seherfig2.eps}}}
%\caption{The order parameters $\Phi_m$ for $\lambda=1$, with $N=100$.  
%There is a set of $N$ effective thresholds, below which hybrid species
%arise, with $m$ sequentially mutated alleles. } %\end{figure}

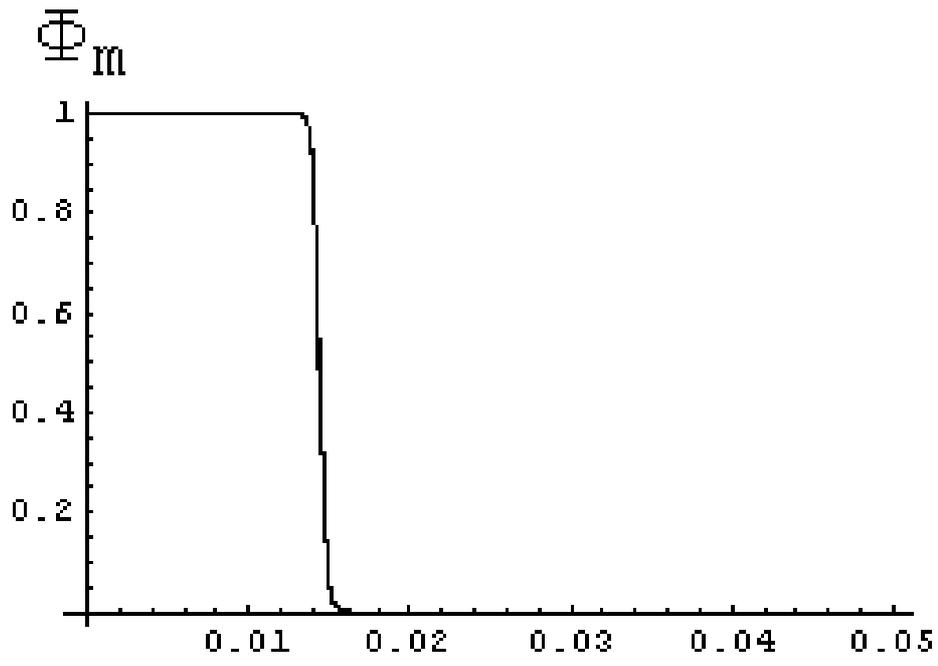
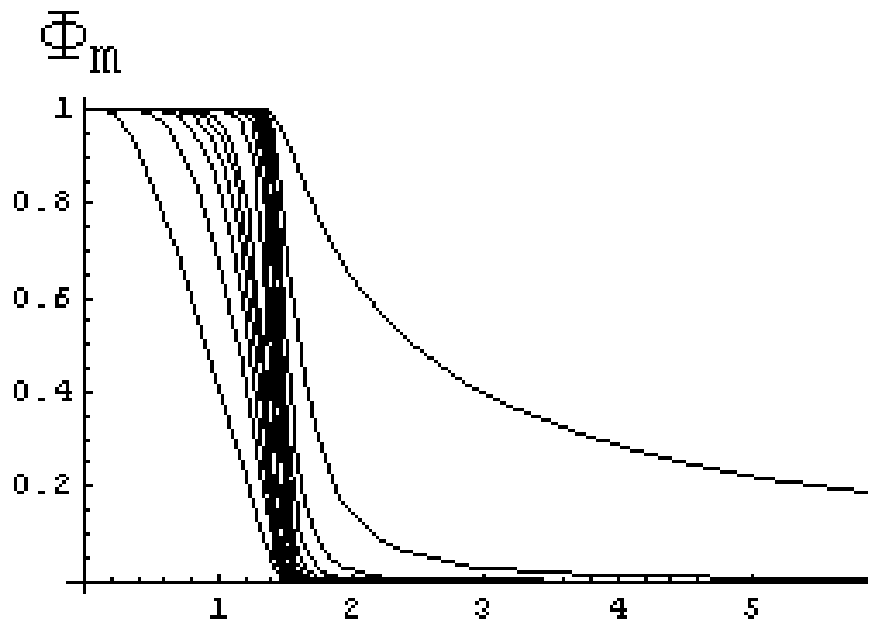
\end{document}